\documentclass[epj]{svjour}

\usepackage{graphics}
\usepackage{epsfig}
\usepackage{amsmath}
\usepackage{amssymb}
\usepackage{color}
\usepackage{tabularx}
\usepackage{enumerate}
\usepackage{graphicx}
\usepackage{float}
\usepackage{xcolor}

\definecolor{model0}{RGB}{255,0,0}
\definecolor{model1}{RGB}{0,255,255}
\definecolor{model2}{RGB}{255,0,255}
\definecolor{model3}{RGB}{255,255,0}
\definecolor{model4}{RGB}{50,50,50}
\definecolor{model5}{RGB}{0,255,0}
\definecolor{model6}{RGB}{160,32,240}
\definecolor{model7}{RGB}{255,165,0}
\definecolor{model8}{RGB}{200,200,200}

\newcommand{\coloredcircle}[1][black]{\Large\textcolor{#1}{\ensuremath\bullet}}

\begin{document}

\title{Effects of single-particle potentials on the level density parameter}

\author{B. Canbula, R. Bulur, D. Canbula \and H. Babacan} 
\institute{Department of Physics, Faculty of Arts and Sciences, Celal Bayar University, 
45140, Muradiye, Manisa, Turkey}

\date{Received: date / Revised version: date}

\mail{bora.canbula@cbu.edu.tr}

\abstract{
The new definition of the energy dependence for the level density 
parameter including collective effects depends strongly on the 
semi-classical approach. For this method, defining an accurate 
single-particle potential is of great importance. The effect of 
the single-particle potential terms, which are central, spin-orbit, 
harmonic oscillator, Woods-Saxon and Coulomb potential, both for 
spherical and deformed cases, on the level 
density parameter was investigated by examining the local success of the 
global parameterizations of eight different combinations of these 
terms. Among these combinations, the sum of the central, spin-orbit, 
harmonic oscillator and Coulomb potentials, gives the most accurate 
predictions compared with experimental data. The local selections 
of the global parameterizations show that the single-particle 
models, which are based on Woods-Saxon potential as the main term, 
are more suitable candidates than the models based on harmonic 
oscillator potential to extrapolate away far from stability. Also 
it can be concluded that the contribution of the Coulomb interaction, 
both around the closed and open shells is not neglectable. 
\PACS{
{21.10.Ma}{Level density} \and 
{21.10.Re}{Collective levels}
     }
}

\authorrunning{B. Canbula et al.}
\maketitle

\section{Introduction}

The number of the excited states in an infinitesimal amount of 
energy around a certain excitation energy is called as the nuclear 
level density (NLD). The NLD is of vital importance for the 
theoretical studies of nuclear structure and reactions. The 
excited levels of the nucleus are very scarce at low excitation 
energy and can be countable easily, but with the increasing 
excitation energy, it is not possible to count the levels 
since the spacing between consecutive levels becomes so narrow. 
Therefore, a function is needed to describe the distribution 
of the excited levels, which is very important for the Hauser-Feshbach 
calculations of the compound-nucleus cross sections.

To develop a theoretical framework for understanding the unusual 
properties of the light exotic beams has been of major interest 
during the last few decades \cite{tanihata1985}. Even though very 
sophisticated nuclear reaction models 
\cite{satchler1983,tamura1965,tobocman1955,rawitscher1974,sakuragi1982,sakuragi1986,yamagata1989,sinha1975,satchler1979,lapoux2008}, 
which give accurate predictions in many cases, have been developed, 
in the solution of this complex problem, structural properties, such 
as the distribution of their excited levels, arising from both pure 
single-particle and collective excitations, should be also considered. 
Thus, correct description of NLD containing these effects is also 
required to explain the reaction data.

A Laplace-like formula for the level density parameter including 
collective effects has been proposed in our recent paper 
\cite{canbula2014}. The new definition of the energy dependence 
for the level density parameter significantly improved the agreement 
between predicted and observed excited energy levels. Furthermore, 
the asymptotic level density parameter, $\tilde{a}$, is redescribed 
in a both more physical and more realistic way. This redescription 
takes into account corrections for both the shell and pairing effects 
in addition to the value obtained from the semi-classical approximation 
analytically or numerically. Using the semi-classical approximation 
\cite{canbula2011} requires a well-defined single-particle potential 
because it directly determines $\tilde{a}$, which is the limit value 
of the level density parameter for the energies above the neutron 
separation energy. Therefore, the single-particle potential almost 
remains the only component to improve the success of the reaction 
calculations that uses the level density as an ingredient. In other 
words, the single-particle potential parameterization is still of 
great importance for the level density.

In our previous work \cite{canbula2014}, we have performed the 
global and the local calculations to obtain $\tilde{a}$. In global 
calculation, we used the single-particle potential consists of 
harmonic oscillator, Coulomb and central potential terms with 
global potential parameters. In contrast, the local calculation 
considers the asymptotic level density parameter as a free parameter 
to be adjusted to the experimental data on the mean resonance 
spacing and discrete level scheme for  each nuclei separately. 
Although, as expected, locally adjusted values of the asymptotic 
level density parameter provide much better agreement with the 
experimental data as compared with the global parameterization, 
disregarding of the global potential parameters is not permissible, 
especially, for the nuclei near the driplines. Since there is not 
enough experimental information on the excited energy levels of 
these nuclei, this situation makes impossible to adjust the 
asymptotic level density parameter locally, and therefore, to rely 
on a global parameterization becomes an obligation rather than a 
choice.

There are nearly 2000 nuclei, which have at least two 
experimentally-known excited energy levels \cite{capote2009}. It 
seems that the global potential which has the highest predictive 
power is harmonic oscillator with Coulomb and central terms for all 
these nuclei over the whole mass range \cite{canbula2014}. However, 
various combinations of potential terms can be more suitable in 
certain mass regions. Moreover, there can be possible correlations 
between the potential choice and some other properties of nuclei, 
not only their mass number. In this manner, it is worth to 
investigate that the behavior of the goodness-of-fit estimators, 
which results from the global parameterizations of different 
single-particle potentials, with respect to some fundamental properties 
as well as the mass number of nuclei. 

The aim of this paper is to investigate the role of the 
single-particle potential in the predictive power of the 
semi-classical level density model for eight different combinations 
of the potential terms by comparing the global parameterizations 
to each other and also by analyzing the local success of them. For 
this purpose, four different combinations are constructed on the 
basis of two main potentials, which are harmonic oscillator and 
Woods-Saxon, with and without Coulomb interaction combined with 
the central and spin-orbit term for all combinations. In addition 
to these potentials, we constructed four more combination to investigate 
the effect of deformation with anisotropic harmonic oscillator 
and deformed Woods-Saxon potentials as main potential terms.

The present paper is organized as follows: In Section 
\ref{sec:theory}, we briefly discuss the method used to 
obtain the asymptotic level density parameter $\tilde{a}$. 
Section \ref{sec:goodnessoffit} contains the definition of 
the goodness-of-fit estimators for phenomenological level 
density models. Then, in Section \ref{sec:spp}, we present the 
single-particle potentials, which we used in this work, and 
discuss our results for 1136 nuclei in Section \ref{sec:rd}. 
Finally, a summary of our model and some concluding remarks 
of this paper are given in Section \ref{sec:conclusions}.

\section{\label{sec:theory}Theory}

Many studies of the nuclear level density have been based on 
the Fermi gas model \cite{bethe1937} in which interactions 
between nucleons are ignored. Therefore, nucleons are assumed 
to occupy equispaced single-particle states arising from an 
average nuclear potential. According to this model, one can 
describe the level density at an excitation energy for a 
certain total angular momentum $J$ and parity $\Pi$ 
\begin{multline}\label{eq:ld}
\rho(U,J,\Pi)= \frac{1}{2} \frac{2J+1}{2\sqrt{2\pi}\sigma^3}\textrm{exp} 
\left [-\frac{(J+\frac{1}{2})^2}{2\sigma^2}  \right ] \times \\\frac{\sqrt{\pi}}
{12}\frac{\textrm{exp}\left [ 2\sqrt{a U} \right ]}{a^{1/4}U^{5/4}}
\end{multline}
where the factor $\frac{1}{2}$ corresponds equiparity. The 
remaining ingredients $a$, $U$ and $\mathrm{\sigma^2}$ represent 
the level density parameter, the effective excitation energy 
and the spin cut-off parameter, respectively. For the Fermi 
gas model, the total level density is described by summing 
Eq. \eqref{eq:ld} over all spins 
\begin{equation}\label{eq:totld}
\rho^{\mathrm{tot}}(E_x)= \frac{1}{\sqrt{2 \pi} \sigma} \frac{\sqrt{\pi}}{12} \frac{exp[2 \sqrt{a U}]}{a^{1/4}U^{5/4}}.
\end{equation}
The collective effects arising from the collective motion of many 
nucleons were not taken into account in the Fermi gas model. 
However, these effects play an important role in populating of the 
excited states. In the later studies 
\cite{hagelund1977,ignatyuk1983}, collective effects have been 
considered as vibrational and rotational effects separately and 
included in the model as additional enhancement factors to total 
level density. In our recent work \cite{canbula2014}, we have 
introduced a Laplace-like formula for the energy dependence of the 
level density parameter which spread the collective effects through 
the whole level density calculation. The level density parameter 
$a$ including collective effects is given by 
\begin{equation}\label{eq:aldp}
a(U) = \tilde{a} \left ( 1 + A_{c} \frac{S_{n}}{U} \frac{\exp (- | U - E_{0} | / {\sigma\prime}_{c}^{3})}{{\sigma\prime}_{c}^{3}} \right ).
\end{equation}
where $S_n$ is the neutron separation energy. $A_c$ is the 
collective amplitude and defined as the shape dependent shell 
(microscopic) correction energy at a critical temperature 
$T_{c}=\sqrt{S_{n}/\tilde{a}}$, which is the nuclear temperature at $S_{n}$, 
\begin{eqnarray}
\label{eq:collectiveamplitude}
A_{c} & = & S(N,Z,T_{c},\mathrm{Shape}) \\ \nonumber
      & = & \left [ M_{\mathrm{exp}}-M_{\mathrm{LDM}} \right ] \frac{\tau_{c}}{\sinh \tau_{c}} \\ \nonumber
	  & = & \left [ M_{\mathrm{exp}}-(M_{0}+E\theta^{2}) \right ] \frac{\tau_{c}}{\sinh \tau_{c}},
\end{eqnarray}
where $\tau_{c}=2\pi^{2}T_{c}/\hbar\omega$. $M_{\mathrm{LDM}}$ 
is the calculated mass of the deformed nucleus from the shape 
dependent liquid drop model, where $M_{0}$ is the mass of the 
corresponding spherical nucleus. Thus, $M_{\mathrm{LDM}}$ 
can be calculated from the formula 
\begin{equation}
\label{eq:liquiddropmass}
M_{\mathrm{LDM}}=M_{N}N+M_{H}Z+E_{V}+E_{S}+E_{C}\pm \frac{11}{\sqrt{A}}+E\theta^{2} 
\end{equation}
of finite-range liquid-drop model \cite{myers1966}. The last 
term is due to small deformations and related to both fissility 
and deformation parameters as 
$E=(2/5)c_{2}A^{2/3}(1-x)\alpha_{0}^{2}$ 
where $\alpha_{0}^{2}=5(a/r_{0})^{2}A^{-2/3}$, 
$\theta=\alpha/\alpha_{0}$, $\alpha^{2}=(5/4\pi)\beta^{2}$, and 
$x=E_{C}/2E_{S}$. The values of the constants of the liquid drop model 
are taken as in Ref. \cite{mengoni1994}. As 
mentioned above, $\sigma_{c}^{2}$ is the spin cut-off parameter 
and is given as
\begin{equation}
\label{eq:criticalspincutoff}
\sigma_{c}^{2}=\frac{T_{c}}{\hbar^{2}} 0.4 M R^{2} 
\left [ 1 + \sqrt{\frac{5\pi}{16}}\beta_{2} 
+ \frac{45\beta_{2}^{2}}{28\pi} 
+ \frac{15\beta_{2}\beta_{4}}{7\sqrt{5}\pi} \right ]
\end{equation}
for deformed nuclei \cite{hagelund1977}. 
Where $c$ indicates to its value at $T_{c}$, 
and ${\sigma\prime}_{c}^{3}=\sigma_{c}^{3}/\tilde{a}$ is the scale 
parameter for Laplace distribution, for further 
details see ref. \cite{canbula2014}. $E_0$ is the energy of 
the first phonon level which corresponds to the first excited 
state caused by vibrational effects \cite{rowe1970,krane1987}. 
This energy level is also known as the first $2^+$ excitation 
state for even-even nuclei \cite{krane1987} and its energy can 
be given with a simple formula $0.2 \hbar \omega$ fitted to  
experimental $2^+$ states \cite{ring1980,siegbahn1965}. Finally, 
$\tilde{a}$ is the asymptotic level density parameter, which is 
crucial in the level density calculations and it is defined in 
different forms by many authors. The simplest expression of this 
parameter is given by \cite{ignatyuk1983,ericson1960} 
\begin{equation}\label{eq:asyp_0}
\tilde{a}=\frac{A}{k}
\end{equation}
or can be taken as the liquid drop like formula \cite{bartel2006}
\begin{multline}\label{eq:asyp_1}
\tilde{a}=a_{\mathrm{vol}} \left [ 1 + k_{\mathrm{vol}} \left (\frac{N-Z}{A} \right )^2 \right ] A + \\
a_{\mathrm{sur}} \left [ 1 + k_{\mathrm{sur}} \left (\frac{N-Z}{A}\right )^2  \right ] A^{2/3} +
a_{\mathrm{Coul}} Z^2 A^{-1/3}.
\end{multline}
Another expression for this parameter, fitting to resonance 
spacings and/or discrete levels, can be written as 
\cite{iljinov1992}
\begin{equation}\label{eq:asyp_2}
\tilde{a}=\alpha A + \beta A^{2/3},
\end{equation}
where $\alpha$ and $\beta$ are adjustable parameters.

Unlike the above expressions, we used a modified expression of the 
well-known semi-classical formula \cite{bohr1998,brack1997} for the 
asymptotic level density parameter in terms of the single-particle 
level density at Fermi energy of nucleus including the shell and 
pairing corrections \cite{canbula2014}
\begin{multline}\label{eq:asyp}
\tilde{a}=\frac{\pi^2}{6}[g_p(E_F^p+S(N,Z)-\Delta) + \\ g_n(E_F^n+S(N,Z)-\Delta) ].
\end{multline} 
$S(N,Z)$ denotes the shell correction energy from the liquid drop 
model \cite{myers1966}. The pairing correction energy is given by 
$\Delta=n\frac{12}{\sqrt{A}}$ with $n$ is $-1$ for odd-odd, $1$ for 
even-even, $0$ for odd nuclei. Therefore, including the energy shift 
$\Delta$ to the Fermi energy substitutes the usage of the expression 
$U=E_x-\Delta$ and allows to use effective excitation energy $U$ 
instead of pure excitation energy $E_x$ directly. $g_p$ and $g_n$ are 
proton and neutron single particle level density, respectively, 
and can be calculated from the semi-classical formula with spin 
degeneracy \cite{brack1997,salasnich2000}
\begin{equation}\label{eq:spld}
g(\varepsilon) = \frac{2}{\pi} \left(\frac{2m}{\hbar^2} \right)^{3/2}
\int r^2 \sqrt{\varepsilon-V(r)} \ \textrm{dr}.
\end{equation} 
$m$ is the mass of a nucleon and $V(r)$ is an effective potential. 
The proton and neutron Fermi energy values $\mathrm{E_F^{\alpha}}$ 
can be obtained by inverting the integral, which gives the nucleon 
number in terms of single-particle level density
\begin{equation}\label{eq:spconservation}
\mathcal{N}_{\alpha} = \int_{- \infty}^{E_{F}^{\alpha}} g_{\alpha}(E) dE, 
\qquad\qquad \mathcal{N}_{\alpha} = \lbrace N,Z \rbrace.
\end{equation} 
Therefore, the crucial role of the single-particle potential in the 
asymptotic level density parameter motivated us to investigate the 
effects of the single-particle potential description to predictive 
power of the semi-classical level density model. For this purpose, 
we consider eight different combinations of various single-particle 
potential terms and analyze the results in the view of agreement 
between their predictions and observations. 

\section{\label{sec:goodnessoffit}Goodness-of-fit Estimators}

Phenomenological level density models have been needed to 
agree with two observable, which are average resonance spacings 
and discrete level schemes. One can test the reliability of the 
level density models with the aid of these observable. In this 
study, we have calculated the rms deviation factor of the mean 
resonance spacings for 289 nuclei, which exist naturally on 
Earth, and their experimental average resonance spacing data 
are available. However, the average goodness-of-fit estimator of 
discrete levels for 1136 nuclei, which have sufficient information 
on the discrete energy level scheme. The goodness-of-fit estimator 
$\mathrm{\chi^2}$ for average resonance spacings has been 
minimized to follow as 
\begin{equation}\label{eq:chi2d}
\chi_{\mathrm{D},i}^{2} = \left( \frac{D_{0,i}^{\mathrm{theo}} -
D_{0,i}^{\mathrm{exp}}}{ D_{0,i}^{\mathrm{err}}} \right
)^{2},
\end{equation}
where the index $i$ indicates the nucleus. 
$\mathrm{D_{0,i}^{\mathrm{exp}}}$ and 
$\mathrm{D_{0,i}^{\mathrm{err}}}$ are respectively experimental 
data and the uncertainty of the average resonance spacing which 
its theoretical predictions are obtained from the equation below 
\begin{equation}\label{eq:d0theo}
\frac{1}{D_{0}^{\mathrm{theo}}} = \sum_{J=\left| I-\frac{1}{2}
\right|}^{J=I+\frac{1}{2}} \rho(S_{n},J,\Pi).
\end{equation}
Unlike the average resonance spacings, goodness-of-fit estimator 
for discrete levels has no experimental error in the cumulative 
level scheme and is given by 
\begin{equation}\label{eq:chi2lev}
\chi_{\mathrm{lev},i}^{2} = \sum_{k=N_{L}^{i}}^{N_{U}^{i}}
\frac{\left[ N_{\mathrm{cum}}^{i}(E_{k}) - k \right]^{2}}{k}.
\end{equation}
Here, $k$ represents the sum over the discrete levels and the 
cumulative number of levels $\mathrm{N_{\mathrm{cum}}}$ up 
to an excitation energy $E$ is calculated from
\begin{equation}\label{eq:nlcum}
N_{\mathrm{cum}}(E) = N_{L} + \int_{E_{L}}^{E}
\rho^{\mathrm{tot}}(E_{x}) dE_{x}.
\end{equation} 
These estimators allow us to test the agreement between our 
predictions and experimental data besides making comparisons 
with the results of the other level density models. The rms 
deviation factor of mean resonance spacings, which is defined 
as for all N nuclides reads 
\begin{equation}\label{eq:frms}
f_{\mathrm{rms}} = \exp \left[ \frac{1}{N} \sum_{i=1}^{N} \left(
\ln \frac{D_{0,i}^{\mathrm{theo}}}{D_{0,i}^{\mathrm{exp}}} \right
)^{2} \right]^{1/2}
\end{equation}
and the average goodness-of-fit estimator for discrete levels is 
\begin{equation}\label{eq:flev}
f_{\mathrm{lev}} = \frac{1}{N} \sum_{i=1}^{N}
\sum_{k=N_{L}^{i}}^{N_{U}^{i}} \frac{\left[
N_{\mathrm{cum}}^{i}(E_{k}) - k \right ]^{2}}{k}.
\end{equation}

\section{\label{sec:spp}Single-Particle Potentials}

In our recent work \cite{canbula2014}, we have calculated the 
nuclear level density parameter by using the semi-classical 
approximation with the single-particle potential consists of 
harmonic oscillator and central potential terms and also 
Coulomb potential for protons. That study \cite{canbula2014} 
leads us to investigate the single-particle potential's role 
of choosing the best agreement for each nuclei in level density 
calculations. In this study, we used various single-particle 
potential terms, which are central, harmonic oscillator, 
Woods-Saxon, Coulomb, and spin-orbit potentials.  We have 
considered eight different combinations 
constructed from these single-particle potential terms: 
\begin{equation}\label{eq:effective}
V(r)=V_{\mathrm{central}}(r) + V_{\mathrm{main}}(r) + V_{\mathrm{Coulomb}}(r) + V_{\mathrm{SO}}.
\end{equation} 
The central potential is taken into account for all potential 
combinations and is given by the equation below
\begin{equation}\label{eq:centrifuge}
V_{\mathrm{central}}(r)=\hbar^2 l(l+1)/2mr^2
\end{equation} 
where $l$ is the angular momentum. Also spin-orbit potential, 
which is very important for the structure of the single-particle 
states near the Fermi surface, is considered in all combinations 
and usually reads \cite{ring1980} 
\begin{equation}\label{eq:usualspinorbit}
V_{\mathrm{SO}}(r)=\lambda \frac{1}{r} \frac{dV}{dr} (\vec{l}\cdot\vec{s}).
\end{equation}
where $\lambda \approx -0.5 \,\mathrm{fm}^2$. However, using 
harmonic oscillator potential leads a constant 
spin-orbit term. Therefore, we prefer here a radius-independent 
form given as \cite{greiner1996}, 
\begin{equation}\label{eq:spinorbit}
V_{\mathrm{SO}}=C \hbar^{2} (\vec{l}\cdot\vec{s})
\end{equation} 
for all combinations in order to ensure consistency. The constant 
$C$ for spin-orbit potential is typically in the range of $-0.3$ 
to $-0.6 \, \mathrm{MeV}/\hbar^{2}$. We take 
$C=-0.3 \, \mathrm{MeV}/\hbar^{2}$ in our calculations, arbitrarily. 

As the main potential, we 
used harmonic oscillator (HO) or Woods-Saxon (WS) potential both 
in their spherical and deformed forms. 
The HO potential, which has a very convenient form for analytical 
calculations, is described as follows 
\begin{equation}\label{eq:harmonicoscillator}
V_{\mathrm{HO}}(r) = \frac{1}{2} m \omega^{2} r^{2} - V_0
\end{equation} 
where $\omega$ is the oscillator frequency, and it has been 
generally parameterized as $41/A^{1/3}$. The depth of the 
potential well, $V_0$, is taken as $50 \,\mathrm{MeV}$ 
\cite{bohr1998}. These parameters are chosen to represent only 
the common properties of the nuclei because the other effects 
like shell and pairing corrections will be applied to Fermi 
energy explicitly as in Eq. \eqref{eq:asyp}. In order to deal 
with the deformation of the nucleus, anisotropic harmonic 
oscillator 
\begin{equation}\label{eq:anisotropicho}
V_{\mathrm{HO}}^{\mathrm{def}}(x,y,z) = \frac{1}{2} m (\omega_{x}^{2} x^{2} + \omega_{y}^{2} y^{2} + \omega_{z}^{2} z^{2}) - V_0
\end{equation} 
is employed. When z-axis is chosen as symmetry axis, one can 
define the oscillator frequencies for axially symmetric shapes 
in terms of the deformation parameter $\delta$ 
\begin{eqnarray}\label{eq:anisotropicfreqs}
\omega_{\perp}^{2}=\omega_{x}^{2}=\omega_{y}^{2} & = & \omega_{0}^{2}(\delta)(1+\frac{2}{3}\delta) \\ \nonumber
\omega_{z}^{2} & = & \omega_{0}^{2}(\delta)(1-\frac{4}{3}\delta).
\end{eqnarray}
$\omega_{0}(\delta)$ is obtained by using the volume conservation 
and reads 
\begin{equation}\label{eq:deformedfreq}
\omega_{0}(\delta)=\omega_{0}{\left[ 1-\frac{4}{3}\delta^{2}-\frac{16}{27}\delta^{3} \right]}^{-1/6}
\end{equation} 
where $\delta$ is related to $\beta$ in Eq. 
\eqref{eq:liquiddropmass} as follows \cite{ring1980}: 
\begin{equation}\label{eq:deltabeta}
\beta\approx\frac{1}{3}\sqrt{\frac{16\pi}{5}}\delta\approx1.057\delta.
\end{equation} 
The other option 
for the main potential is the WS potential, which is a more 
realistic description compared to HO meanwhile to use WS 
in calculations causes some difficulties due to the inability 
to obtain analytical solutions. The WS potential is written as 
\begin{equation}\label{eq:woodssaxon}
V_{\mathrm{WS}}(r) = - \frac{V_0}{1+\mathrm{exp \left( \frac{r-R}{a} \right)}}.
\end{equation}
$V_0$, $R$ and $a$ are depth, radius and diffuseness parameters 
of the potential well, respectively. In this study, potential well 
depth is taken as $50 \,\mathrm{MeV}$ to be consistent with the HO 
potential, and radius is defined as $R = r_0 A^{1/3}$ where $r_0$ 
equals $1.25 \,\mathrm{fm}$. Also, diffuseness parameter of the 
potential well is used as $0.5 \,\mathrm{fm}$ in calculations. For 
WS potential, the deformation can be included in the parameter 
$R$ as 
\begin{equation}\label{eq:deformedradius}
R(\theta,\phi)=(r_0 A^{1/3}) \left[ 1 + \sum_{\lambda}\sum_{\mu} a_{\lambda\mu} Y_{\lambda\mu}(\theta,\phi) \right],
\end{equation}
related to deformation parameter as 
$\beta^{2}=\sum_{\lambda}\sum_{\mu} |a_{\lambda\mu}|^{2}$ 
\cite{myers1966}. Therefore, WS potential for the deformed nuclei 
is expressed as follows:
\begin{equation}\label{eq:deformedws}
V_{\mathrm{WS}}^{\mathrm{def}}(r,\theta,\phi) = - \frac{V_0}{1+\mathrm{exp \left( \frac{r-R(\theta,\phi)}{a} \right)}}.
\end{equation}

Furthermore, the potential combination consisting of the central 
and spin-orbit potential terms and the selected main potential is 
complemented with Coulomb potential for protons to investigate 
Coulomb interaction on the level density parameter. Under the 
assumption that nucleus is a uniformly charged sphere, the Coulomb 
potential is given by 
\begin{equation}\label{eq:coulomb}
V_{C}(r) = \left\lbrace
\begin{array}{ll}
\displaystyle{\frac{Z \mathrm{e}^{2}}{2 R_{C}} \left ( 3 - 
\frac{r^{2}}{{R_{C}}^{2}} \right )} & r \le R_{C} \\
\displaystyle{\frac{Z \mathrm{e}^{2}}{r}} &  r \ge R_{C}
\end{array}
\right.
\end{equation}
where $\mathrm{R_C}$ is charge radius and is taken as 
$R_C = 1.169 A^{0.291}$ from a recent fit \cite{bayram2013} 
to the latest nuclear charge radii data \cite{angeli2013}. 

\begin{table*}[t]
\caption{\label{tab:goodnessoffit} (Color online) Goodness-of-fit 
estimator values for various potential combinations and comparison 
with previous results.}
\centering
\begin{tabularx}{\textwidth}{llXlll}
\hline
\hline
\textrm{Model}&&\textrm{Potential Terms}&$\mathrm{f_{\mathrm{rms}}}$&$\mathrm{f_{\mathrm{lev}}}$&\textrm{Reference}\\
\hline
Model 0 &\coloredcircle[model0]& Local Selections                     & 1.30 & 1.00 & This work\\
Model 1 &\coloredcircle[model1]& Central + SO + HO                    & 1.63 & 1.91 & This work\\
Model 2 &\coloredcircle[model2]& Central + SO + HO + Coulomb          & 1.51 & 1.32 & This work\\
Model 3 &\coloredcircle[model3]& Central + SO + WS                    & 2.06 & 1.65 & This work\\
Model 4 &\coloredcircle[model4]& Central + SO + WS + Coulomb          & 2.11 & 1.70 & This work\\
Model 5 &\coloredcircle[model5]& Central + SO + Deformed HO           & 1.66 & 1.71 & This work\\
Model 6 &\coloredcircle[model6]& Central + SO + Deformed HO + Coulomb & 1.55 & 1.31 & This work\\
Model 7 &\coloredcircle[model7]& Central + SO + Deformed WS           & 1.74 & 1.62 & This work\\
Model 8 &\coloredcircle[model8]& Central + SO + Deformed WS + Coulomb & 2.24 & 1.80 & This work\\
        & & Central + HO + Coulomb & 1.53 & 1.32 & \cite{canbula2014}\\
        & & HO & 1.12 & 43.9 & \cite{canbula2011}\\
        & & HO + Coulomb & 1.16 & 42.6 & \cite{canbula2011}\\
\hline
\end{tabularx}
\end{table*} 

However, one would note that the first single-particle level has 
never been in the bottom of the potential well, and this situation 
should be considered when calculating the Fermi energy level. 
Therefore, the contribution from the interval between the bottom of 
the potential well and the first single-particle level with the 
lowest energy should be zero in the integral 
\eqref{eq:spconservation}. The value of the first single-particle 
level is well known for the HO potential but to make a similar 
prediction for WS potential might be difficult. So, considering 
that the total depth of the well is approximately equal to the 
sum of the Fermi energy and binding energy \cite{bohr1998}, to 
determine a value for this interval is a reasonable correction 
which has applied as first $8 \,\mathrm{MeV}$ for the WS potential. 

\section{\label{sec:rd}Results and Discussion}

Using the single-particle potential terms given in 
Section \ref{sec:spp}, we define eight different combinations, which 
are labelled as Model 1 to 8. Table \ref{tab:goodnessoffit} includes 
these model definitions with their $f_{\mathrm{rms}}$ and 
$f_{\mathrm{lev}}$ values, which are calculated for 1136 nuclei using 
Eqs. \eqref{eq:frms} and \eqref{eq:flev}, respectively, and 
also shows the results of our preceding 
studies \cite{canbula2014,canbula2011}. The unique difference between 
Model 2 and the single-particle potential used in Ref. \cite{canbula2014} 
is the spin-orbit term, which improves the predictive power of the model. 
When this model compared our another study \cite{canbula2011}, two 
remarkable differences exist in model definitions. In this study, we 
replace the simple description ($R_{C}=1.2\,A^{1/3}$) of the charge 
radius with an expression obtained in a recent fit \cite{bayram2013}, 
and also use Laplace-like formula instead of well-known Ignatyuk's 
formula \cite{ignatyuk1975} for level density parameter. These modifications 
lead to a significant improvement in the agreement between the predicted 
cumulative number of levels and the observed number of excited levels. 

\begin{table*}[t!]
\caption{\label{tab:isotopecounter} Details of the selections of 
Model 0. The first line represents 289 stable isotopes. The third 
line represents 1136 isotopes while the selections of the remaining 
847 isotopes are given in the second line.}
\centering
\begin{tabularx}{\textwidth}{|rX|rr|rr|rr|rr|rr|rr|rr|rr|}
\hline
\hline
\textrm{NoI}&&
\textrm{NoI}&\textrm{\%}&
\textrm{NoI}&\textrm{\%}&
\textrm{NoI}&\textrm{\%}&
\textrm{NoI}&\textrm{\%}&
\textrm{NoI}&\textrm{\%}&
\textrm{NoI}&\textrm{\%}&
\textrm{NoI}&\textrm{\%}&
\textrm{NoI}&\textrm{\%}\\
\hline
\multicolumn{2}{l|}{} & 
\multicolumn{2}{c|}{Model 1} & 
\multicolumn{2}{c|}{Model 2} & 
\multicolumn{2}{c|}{Model 3} & 
\multicolumn{2}{c|}{Model 4} & 
\multicolumn{2}{c|}{Model 5} & 
\multicolumn{2}{c|}{Model 6} & 
\multicolumn{2}{c|}{Model 7} & 
\multicolumn{2}{c|}{Model 8} \\
\hline
 289&&  37&12.8&  67&23.1&  30&10.4&  34&11.8&  19& 6.6&  40&13.8&  32&11.1&  30&10.4 \\
 847&&  96&11.3&  61& 7.2& 245&28.9& 130&15.3&  26& 3.1&  38& 4.5&  82& 9.7& 169&20.0 \\
1136&& 133&11.7& 128&11.3& 275&24.2& 164&14.4&  45& 4.0&  78& 6.9& 114&10.0& 199&17.5 \\
\hline
\multicolumn{2}{l|}{} & 
\multicolumn{2}{c|}{Model 1+2} & 
\multicolumn{2}{c|}{Model 3+4} & 
\multicolumn{2}{c|}{Model 5+6} & 
\multicolumn{2}{c|}{Model 7+8} & 
\multicolumn{2}{c|}{Model 1+5} & 
\multicolumn{2}{c|}{Model 2+6} & 
\multicolumn{2}{c|}{Model 3+7} & 
\multicolumn{2}{c|}{Model 4+8} \\
\hline
 289&& 104&35.9&  64&22.2&  59&20.4&  62&21.5&  56&19.3& 107&37.0&  62&21.5&  64&22.2 \\
 847&& 157&18.5& 375&44.2&  64& 7.6& 251&29.7& 122&14.4&  99&11.7& 327&38.6& 299&35.3 \\
1136&& 261&23.0& 439&38.6& 123&10.9& 313&27.5& 178&15.7& 206&18.2& 389&34.2& 363&31.9 \\
\hline
\multicolumn{2}{l|}{} & 
\multicolumn{4}{r|}{Model 1+2+5+6} & 
\multicolumn{4}{r|}{Model 3+4+7+8} & 
\multicolumn{4}{r|}{Model 1+2+3+4} & 
\multicolumn{4}{r|}{Model 5+6+7+8} \\
\hline
 289&&    &    & 163&56.4&    &    & 126&43.6&    &    & 168&58.1&    &    & 121&41.9 \\
 847&&    &    & 221&26.1&    &    & 626&73.9&    &    & 532&62.7&    &    & 315&37.3 \\
1136&&    &    & 384&33.9&    &    & 752&66.1&    &    & 700&61.6&    &    & 436&38.4 \\
\hline
\multicolumn{18}{l}{\begin{footnotesize}NoI stands for number of isotopes.\end{footnotesize}} \\
\end{tabularx}
\end{table*}

\begin{figure}[h!]
\centering
\includegraphics{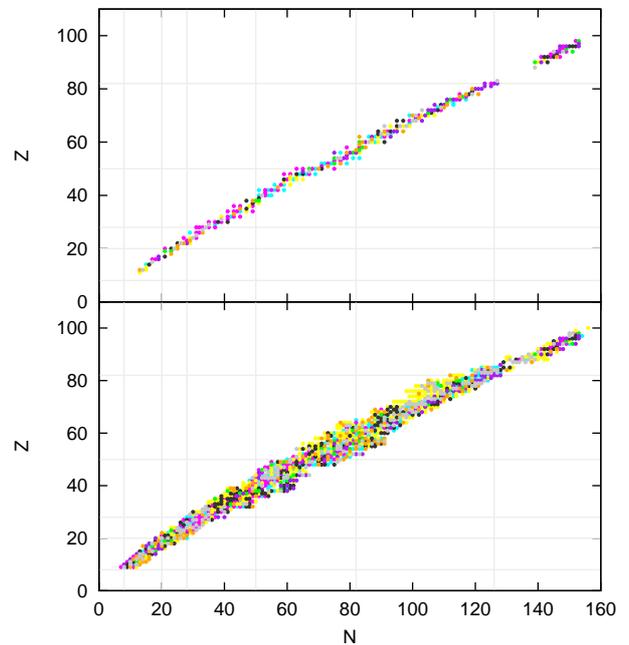}
\caption{\label{fig:fig1} (Color online) 
The local selections of the single-particle combinations. The upper 
panel includes 289 stable isotopes while the lower panel includes 
1136 isotopes. Cyan, yellow, magenta, black, green, purple, orange and 
gray colored dots donate Models 1 to 8, respectively. The magic numbers 
are shown with grid lines.}
\end{figure}

As the model predictions are compared with each other, Model 2 gives 
the most accurate model for the entire mass range. Moreover, 
it can be seen that HO based models are more predictive than WS based 
models, but among them only Model 7, which takes into account the 
deformation of the nucleus, is comparable with HO based models. It 
is an unexpected result because WS potential is more realistic than HO 
potential. This fact led us to a further analysis. On the other hand the 
comparison between these eight models only gives a general idea about 
their success to describe the common properties of the most of the nuclei. 
But in the case of the extrapolation to certain mass regions, especially 
near the driplines, this point of view becomes deficient. Therefore, 
testing the predictions of the global potential parameterizations 
locally for each nuclei might be the only useful method to conclude that 
which potential model is suitable to extrapolate outside the certain 
mass regions. 

Considering these two aspects, we define Model 0, which 
consists of the local selections among the global parameterizations 
of Models 1-8. In Model 0, the selections have been made to give the 
lowest $\chi_{i}^{2}$ contribution to global $\chi^{2}$ values for 
the considered nucleus. Local model selections chosen with this 
criterion are illustrated in the upper and lower panels of Figure 
\ref{fig:fig1} for 289 and 1136 nuclei, respectively. The local model 
selections of 289 stable isotopes of the total of 1136 isotopes are 
shown explicitly in the upper panel because of the interpretations 
about the stable isotopes might be completely different from the rest. 
Consistent with the above discussion of Table \ref{tab:goodnessoffit}, 
for stable isotopes, most of the selections are in HO based models. 
Models 2 and 6 are the almost only options in the heavy mass, $Z>70$, 
region. Therefore, it can be said that the effect of the Coulomb 
interaction becomes indispensable with the increasing proton number, 
specially in this region. Also in $Z<40$ region, Model 2 selections 
are quite intense compared to other models. However, in the case of 1136 
isotopes, a significant increase has been noted in the number of 
selections of WS based models. These selections are notably more 
abundant in the region covering the exterior side of the island shown 
in the lower panel of Figure \ref{fig:fig1}. 

\begin{figure}[h!]
\centering
\includegraphics{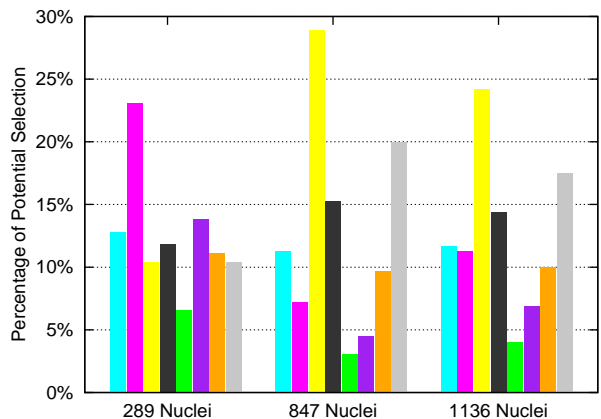}
\caption{\label{fig:fig2} (Color online) 
The bar chart of the selections of Model 0. The color codes are 
the same as Figure \ref{fig:fig1} and Table \ref{tab:goodnessoffit}.}
\end{figure}

On the other hand, quantifying the total numbers of selections of 
models might be useful to conclude that which model is better to 
extrapolate to far from stability. Table \ref{tab:isotopecounter} 
shows this quantification. Besides Models 1-8, the cumulative number 
of selections of models based on the same kind of potential terms are 
also shown in Table \ref{tab:isotopecounter}. The local selections of 
289 stable isotopes, and the remaining 847 isotopes seem to be completely 
different from each other. This issue is also emphasized in Figure 
\ref{fig:fig2}. For 289 stable isotopes WS based models have been 
selected only 43.6 percent of total selections. However, for remaining 
847 isotopes, selection rate increases to 73.9 percent. In contrast to 
overwhelming superiority of Model 2, which is clearly seen from Table 
\ref{tab:goodnessoffit}, WS based models, especially Model 3, seem to 
have had considerable success in describing the properties of the isotopes 
far from stability. Moreover, in the case of 1136 isotopes, the total 
numbers of selections of WS based models are slightly greater than that 
of HO based models. When the selections are expanded from 289 isotopes 
to 1136, Model 2, which is the most selected model, falls back to 
fifth place. Therefore, it can be concluded that HO potential is very 
suitable for describing the most of the nuclei, but some certain 
isotopes or mass regions can be described better with WS potential. 
Nevertheless, WS potential is not suitable for a generalization to 
the entire mass region, at least with the potential parameters used 
in this work. 

\begin{figure}[h!]
\centering
\includegraphics{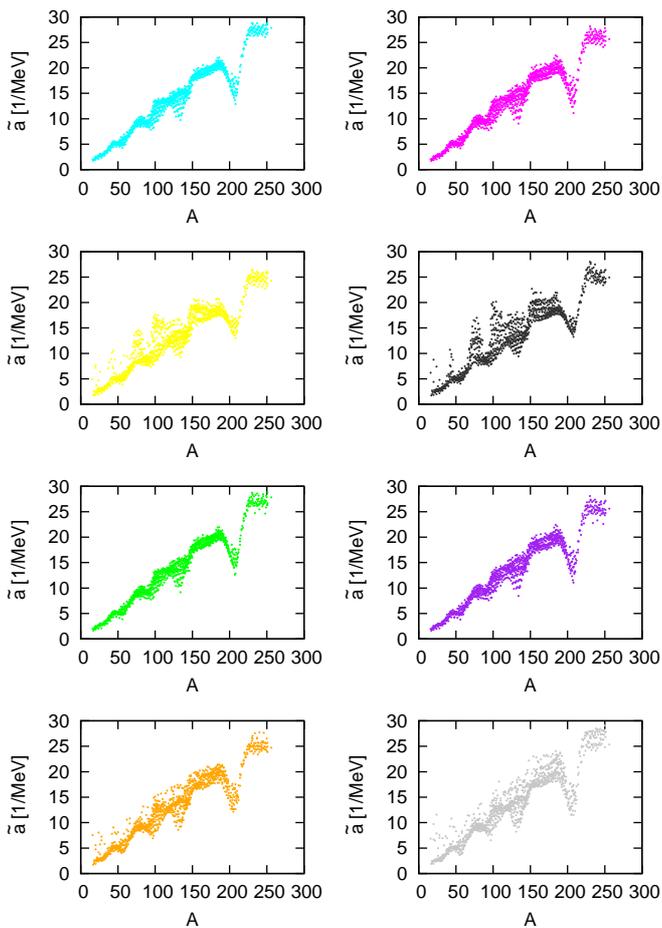}
\caption{\label{fig:fig3} (Color online) 
The asymptotic level density parameter values obtained by using eight 
different single particle potential combinations for 1136 isotopes. 
Models 1 and 2 are given in the first row, while Models 3 and 4 are 
given in the second row, and Models 5 and 6 are given in the third row 
and finally Models 7 and 8 are given in the forth row from left to 
right, respectively. The color codes are the same as Figure \ref{fig:fig1} 
and Table \ref{tab:goodnessoffit}.}
\end{figure}

The asymptotic level density parameter values obtained by using 
Models 1-8 are illustrated in Figure \ref{fig:fig3}. The first 
impression that emerges from Figure \ref{fig:fig3}, for all models there 
exists the downwards peaks around the closed shells arising from the shell 
effects. However, there are also upwards peaks around the open shells only 
for Models 3 and 4 (WS based spherical models). In addition, the 
Coulomb term seems to increase the depth and the height of these peaks. 

\begin{figure}[h!]
\centering
\includegraphics{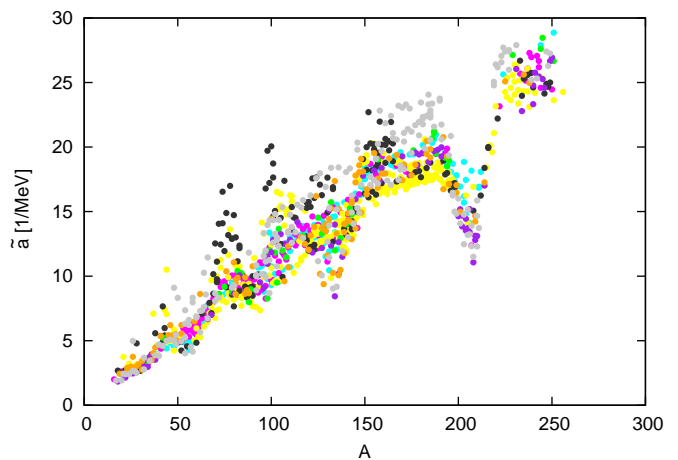}
\caption{\label{fig:fig4} (Color online) 
The asymptotic level density parameter values obtained by using 
Model 0, which is the local selections of the global 
parameterizations of Models 1-8. The color codes are 
the same as Figure \ref{fig:fig1} and Table \ref{tab:goodnessoffit}.} 
\end{figure}

In Figure \ref{fig:fig4} the asymptotic level density parameter values 
of Model 0 are plotted. The color codes correspond to the single-particle 
potential models used for calculating the asymptotic level density 
parameter. Models based on HO potential are selected only for 
the isotopes which are weakly influenced by the shell effects. This 
situation is also consistent with the issue which is mentioned above 
in the discussion of Figure \ref{fig:fig1} as well as Table 
\ref{tab:isotopecounter}. However, Model 6 has been selected around 
the closed shell near the mass number $A$ equal to 208, this 
contradiction is caused by the fact that the demand to select the 
model which gives the deepest peak in this region. Similarly, around 
the open shells, the selections have been a model including Coulomb 
term but this time it is Model 4 or 8, which gives the highest peak in 
these regions. Therefore, it can be concluded from Figure \ref{fig:fig4} 
that it is essential to consider Coulomb interaction for the isotopes 
around both closed and open shells.

\begin{figure}[h!]
\centering
\includegraphics{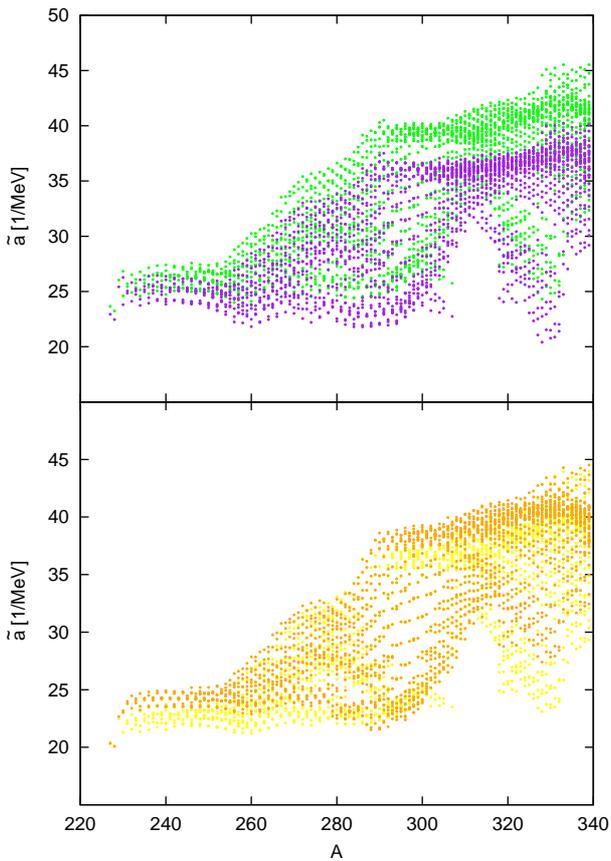}
\caption{\label{fig:fig5} (Color online) 
The asymptotic level density parameter values for superheavy nuclei, 
which have atomic number $Z>100$. The upper and lower panels represent 
the comparisons between Models 5 and 6, and Models 3 and 7, respectively. 
The color codes are the same as Figure \ref{fig:fig1} and Table \ref{tab:goodnessoffit}.}
\end{figure}

Finally, Figure \ref{fig:fig5} represents the asymptotic level density 
parameter values for nuclei, which are known as superheavy elements, 
in the mass region $Z>100$. The upper panel includes the comparison 
between Models 5 and 6 to discuss the effect of Coulomb term in this 
mass region. It can be clearly seen that Coulomb term has an inhibitory 
effect on the asymptotic level density parameter for the increasing 
proton numbers. To discuss the effect of the deformation, the comparison 
between Models 3 and 7 is given in the lower panel. The values resulting 
from Model 7, which is the deformed version of Model 3, are slightly 
greater than the predictions of Model 3. On the other hand, one can easily 
say that the asymptotic level density parameter tends to reach a limit 
value with the increasing mass number. 

\begin{figure}[h!]
\centering
\includegraphics{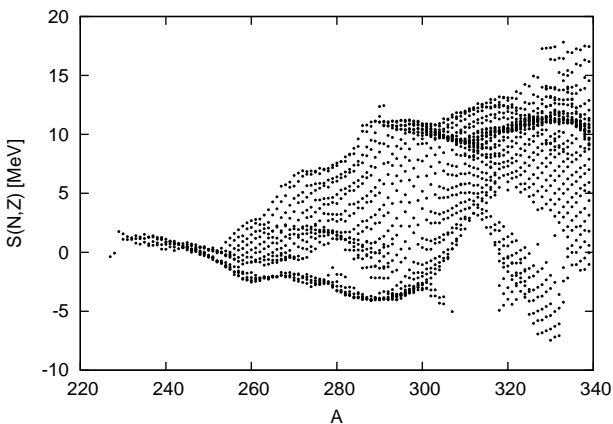}
\caption{\label{fig:fig6} 
The shell correction energies for superheavy nuclei. The liquid drop 
model parameters are taken as in Ref. \cite{mengoni1994}.}
\end{figure}

To extract the asymptotic value of the level 
density parameter for a certain superheavy  nucleus, one can easily use 
the conservation condition \eqref{eq:spconservation} with the selected 
single particle potential and compute the Fermi energy. Thus, asymptotic 
level density parameter can be calculated by using Eq. \eqref{eq:asyp}, 
which also includes the shell and pairing corrections. While the pairing 
correction energy can be calculated from its traditional definition (see 
the text after Eq. \eqref{eq:asyp}), the calculated shell correction 
energies with the liquid drop model \cite{myers1966} by using the parameter 
values as in Ref. \cite{mengoni1994} are given in Figure \ref{fig:fig6}. 
As expected, a strong correlation between $\tilde{a}$ and $S(N,Z)$ values 
is observed from Figs. \ref{fig:fig5} and \ref{fig:fig6}. Therefore, a 
reliable extrapolation to the nuclei far from stability requires both 
a well-defined single-particle potential and a parameter set for liquid 
drop model. Because of the fact that the energy dependence of the level 
density parameter is completely defined by the collective amplitude 
$A_{c}$, deformation parameters are required for both obtaining 
$a(U)$ and an asymptotic value of the level density parameter with 
deformed single-particle models. The calculated values of the deformation 
parameters with Finite Range Droplet Model \cite{moller1995} are used 
for nuclei where experimental information is not available.

\section{\label{sec:conclusions}Conclusions}

Summarizing, the effect of the single-particle potential terms, which 
are central, spin-orbit, harmonic oscillator, Woods-Saxon and 
Coulomb potential, both for spherical and deformed cases, 
on the level density parameter was investigated by examining the 
local success of the global parameterizations of eight different 
combinations of these terms. In the light of above discussions, the 
following conclusions can be drawn from this study: 

\begin{enumerate}[(i)]
\item Model 2, which is the sum of the central, spin-orbit, harmonic 
oscillator and Coulomb potentials, gives the most accurate predictions 
compared to experimental data. 
\item The local selections of the global parameterizations indicate 
that the single-particle models, which are based on Woods-Saxon 
potential as the main term, are more suitable candidates than the 
models based on harmonic oscillator potential to extrapolate away far 
from stability. 
\item It is seen from the investigation of the asymptotic level 
density parameters obtained from the local selection that the 
contribution of Coulomb interaction is not ignorable both around 
the closed and open shells. 
\item Finally, for the exotic and superheavy nuclei, which have not any 
experimental information to adjust the level density parameters, 
the single-particle potential consists of the central, spin-orbit, 
Woods-Saxon, and Coulomb potential terms is the most reliable 
potential model to calculate the asymptotic level density parameter. 
\end{enumerate}

\section*{Acknowledgements}

This work was supported by the Turkish Science and Research 
Council (T\"{U}B\.{I}TAK) under Grant No. 112T566. Bora Canbula 
acknowledges the support through T\"{U}B\.{I}TAK PhD Program 
fellowship B\.{I}DEB-2211 Grant.


\begin{thebibliography}{34}
\bibitem{tanihata1985} I. Tanihata, H. Hamakagi, O. Hashimoto, Y. Shida, N. Yoshikawa, K. Sugimoto, O. Yamakawa, T. Kobayashi, N. Takahashi, Phys. Rev. Lett. \textbf{55}, 2676-2679 (1985).
\bibitem{satchler1983} G.R. Satchler, \textit{Direct Nuclear Reactions}, Clarendon Press, Oxford (1983).
\bibitem{tamura1965} T. Tamura, Rev. Modern Phys. \textbf{37}, 679-708 (1965).
\bibitem{tobocman1955} W. Tobocman, M.H. Kalos, Phys. Rev. \textbf{97}, 132-136 (1955).
\bibitem{rawitscher1974} G.H. Rawitscher, Phys. Rev. C \textbf{9}, 2210-2229 (1974).
\bibitem{sakuragi1982} Y. Sakuragi, M. Yahiro, M. Kamimura, Prog. Theoret. Phys. \textbf{68}, 322-326 (1982).
\bibitem{sakuragi1986} Y. Sakuragi, Y. Masanobu, K. Masayasu, Prog. Theo. Phys. Suppl. \textbf{89}, 136-211 (1986).
\bibitem{yamagata1989} Y. Yamagata, K. Yuasa, N. Inabe, M. Nakamura, M. Tanaka, S. Nakayama, K. Katori, M. Inoue, S. Kubono, T. Itahashi, H. Ogata, Y. Sakuragi, Phys. Rev. C \textbf{39}, 873-876 (1989).
\bibitem{sinha1975} B. Sinha, Phys. Rep. \textbf{1}, 1-57 (1975).
\bibitem{satchler1979} G.R. Satchler, W.G. Love, Phys. Rep. \textbf{3}, 183-254 (1979).
\bibitem{lapoux2008} V. Lapoux et al., Phys. Lett. B \textbf{658}, 198-202 (2008).
\bibitem{canbula2014} B. Canbula, R. Bulur, D. Canbula, H. Babacan, Nuclear Physics A \textbf{929}, 54-70 (2014).
\bibitem{canbula2011} B. Canbula, H. Babacan, Nucl. Phys. A \textbf{858}, 32-47 (2011).
\bibitem{capote2009} R. Capote et al., Nucl. Data Sheets \textbf{110}, 3107-3214 (2009).
\bibitem{bethe1937} H.A. Bethe, Rev. Mod. Phys. \textbf{9}, 69 (1937).
\bibitem{hagelund1977} H. Hagelund, A.S. Jensen, Phys. Scr. \textbf{15}, 225-236 (1977).
\bibitem{ignatyuk1983} A.V. Ignatyuk, \textit{The Statical Properties of the Excited Atomic Nuclei}, Energoatomizdat, Moscow (1983).
\bibitem{myers1966} W.D. Myers, W.J. Swiatecki, Nucl. Phys. \textbf{81}, 1-60 (1966).
\bibitem{mengoni1994} A. Mengoni, Y. Nakajima, J. Nucl. Sci. Tech. \textbf{31}, 151-162 (1994).
\bibitem{rowe1970} D.J. Rowe, \textit{Nuclear Collective Motion}, Methuen, London (1970).
\bibitem{krane1987} K.S. Krane, \textit{Introductory Nuclear Physics}, John Wiley and Sons Inc. (1987).
\bibitem{ring1980} P. Ring, P. Schuck, \textit{The Nuclear Many-Body Problem}, Springer (1980).
\bibitem{siegbahn1965} K. Siegbahn, \textit{Alpha-, Beta- and Gamma-Ray Spectroscopy}, North-Holland (1965).
\bibitem{ericson1960} T. Ericson, Adv. Phys. \textbf{9} , 425-511 (1960).
\bibitem{bartel2006} J. Bartel, K. Pomorski, B. Nerlo-Pomorska, Int. J. Mod. Phys. E \textbf{15}, 478-483 (2006).
\bibitem{iljinov1992} A.S. Iljinov, M.V. Mebel, N. Bianchi, E. De Sanctis, C. Guaraldo, V. Lucherini, V. Muccifora, E. Polli, A.R. Reolon, P. Rossi, Nucl. Phys. A \textbf{543}, 517-557 (1992).
\bibitem{bohr1998} A. Bohr, B.R. Mottelson, \textit{Nuclear Structure}, W. A. Benjamin, Inc. (1998).
\bibitem{brack1997} M. Brack, R.K. Bhaduri, \textit{Semiclassical Physics}, Addison-Wesley Publishing Company, Inc. (1997).
\bibitem{salasnich2000} L. Salasnich, J. Math. Phys. \textbf{41}, 8016-8024 (2000).
\bibitem{greiner1996} W. Greiner, J.A. Maruhn, \textit{Nuclear Models}, Springer-Verlag (1996).
\bibitem{bayram2013} T. Bayram, S. Akkoyun, S.O.Kara,A. Sinan, Acta Phys. Pol. B \textbf{44}, 1971-1799, (2013).
\bibitem{angeli2013} I. Angeli, K.P. Marinova, At. Nucl. Data Tables \textbf{99}, 69-95 (2013).
\bibitem{ignatyuk1975} A.V. Ignatyuk, G.N. Smirenkin, A.S. Tishin, Sov. J. Nucl. Phys. \textbf{21}, 255 (1975).
\bibitem{moller1995} P. Moller, J.R. Nix, W.D. Myers, W.J. Swiatecki, At. Nucl. Data Tables \textbf{59}, 185 (1995); P. Moller, J.R. Nix, At. Nucl. Data Tables \textbf{26}, 165 (1981); G. Audi, A.H. Wapstra, C. Thibault, Nucl. Phys. A \textbf{729}, 337 (2003).
\end{thebibliography}
\end{document}